\documentclass[prl,twocolumn, showpacs,preprintnumbers,amsmath,amssymb]{revtex4}

\usepackage{graphicx,color}
\usepackage{multirow}

\begin{document}

\title{Near-unity Cooper pair splitting efficiency}

\author{J.~Schindele}
\author{A.~Baumgartner}
\email{andreas.baumgartner@unibas.ch}
\author{C.~Sch\"{o}nenberger}
\affiliation{Department of Physics, University of Basel, Klingelbergstrasse 82, CH-4056 Basel, Switzerland}

\date{\today}

\begin{abstract}
The two electrons of a Cooper pair in a conventional superconductor form a singlet and therefore a maximally entangled state. Recently, it was demonstrated that the two particles can be extracted from the superconductor into two spatially separated contacts via two quantum dots (QDs) in a process called Cooper pair splitting (CPS). Competing transport processes, however, limit the efficiency of this process. Here we demonstrate efficiencies up to $90$\%, significantly larger than required to demonstrate interaction-dominated CPS, and on the right order to test Bell's inequality with electrons. We compare the CPS currents through both QDs, for which large apparent discrepancies are possible. The latter we explain intuitively and in a semi-classical master equation model. Large efficiencies are required to detect electron entanglement and for prospective electronics-based quantum information technologies.
\end{abstract}


\maketitle

Quantum entanglement between two particles is a fundamental resource for quantum information technologies \cite{Vidal_PRL91_2003}. Experiments with entangled photons are well developed and already offer first applications \cite{Ursin_Zeilinger_NaturePhys3_2007}. However, entanglement between electrons, the fundamental particles of electronics, is difficult to create in a controlled way. Electron-electron interactions, for example in the Fermi sea of a metal, tend to destroy particle correlations. In contrast, in nanostructured electronic devices interactions can be exploited, for example to generate entangled photons \cite{Salter_Nature465_2010}.

An elegant source of electronic entanglement are singlet two-particle ground states, for example in the naturally occurring BCS state of a conventional superconductor. It was proposed to use such a superconductor to produce spatially separated entangled electrons in a process known as crossed Andreev reflection or Cooper pair splitting (CPS) \cite{Recher_Loss_PRB63_2001, lesovik_martin_blatter_EPJB01}. Though metallic structures show electronic correlations due to superconductivity \cite{Beckmann_PRL93_2004, Russo_Klapwijk_PRL95_2005, Cadden-Zimansky_Chandrasekhar_NatPhys_2009, Kleine_Baumgartner_EPL87_2009, Wei_Chandrasekhar_NatPhys_2010}, their tunability is minimal. Recently, CPS was demonstrated on devices where a superconducting contact is coupled to two parallel quantum dots (QDs), each with a normal metal output lead, as shown schematically in Fig.~1a \cite{Hofstetter2009, Herrmann_Kontos_Strunk_PRL104_2010, Hofstetter_Baumgartner_PRL107_2011, Das_Heiblum_private_comm}.
In the latter experiments the CPS efficiency ranged from a few percent up to 50\%. Such values can in principle be reached without electron-electron interactions, e.g. in a chaotic cavity \cite{Samuelsson_Buettiker_PRB66_2002}, or in a double-dot system with strong inter-dot coupling \cite{Herrmann_Kontos_Strunk_PRL104_2010}, where the electrons of a Cooper pair can exit the device through two ports at random. However, for applications and more sophisticated experiments, for example the explicit demonstration of entanglement, efficiencies close to unity are required.

Here we present CPS experiments on a carbon nanotube (CNT) device and demonstrate efficiencies up to 90\%, values only possible with electron-electron interactions. In addition, we find discrepancies when extracting the CPS part of the currents through the two quantum dots, which we relate to a competition between local processes and CPS in a semi-classical master equation model. The large CPS efficiencies and the increased understanding of the relevant mechanisms are important steps on the way to an all-electronic source of entangled electron pairs in a solid-state device.

\begin{figure}[b]
\centering
\includegraphics{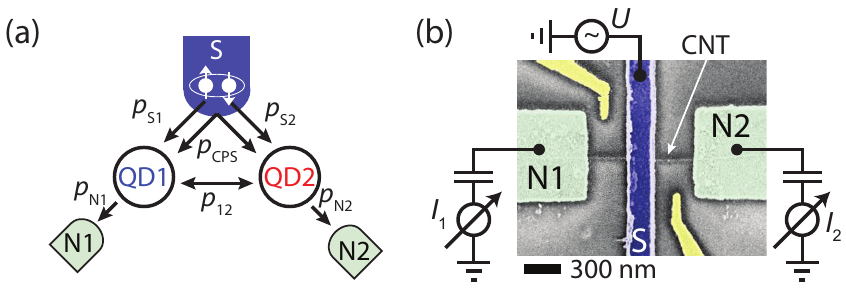}
\caption{(Color online) (a) Schematic of the device and relevant process probabilities. (b) Scanning electron micrograph of the CPS device and measurement setup.}
\end{figure}

An artificially colored scanning electron micrograph of a CPS device is shown in Fig.~1b, together with a schematic of the measurement. A CVD-grown CNT (arrow) is contacted in the center by an aluminum contact (S), which becomes superconducting below $\sim 1.1\,$K and is evaporated on a $4\,$nm palladium (Pd) contact layer. Two Pd contacts to the right and left of S serve as normal metal contacts N1 and N2, both of which define a quantum dot (QD1 and QD2) on the two CNT segments adjacent to S. The QDs can be tuned electrically by a global backgate and the local side-gates SG1 and SG2.

From standard charge stability diagrams we extract charging energies of $\sim7\,$meV for QD1 and $\sim4\,$meV for QD2, an orbital energy spacing of $\sim1\,$meV, and an energy gap due to the superconductor of $\sim 120\,\mu$eV. With S in the normal state we find typical level broadenings of $\sim150$-$500\,\mu$eV. Relatively low peak conductances suggest rather asymmetric coupling of the QDs to the leads. The lever arms from a side-gate across the superconductor to the other QD is roughly ten times smaller than that of a local side gate. In conductance measurements with the two QDs in series we do {\it not} observe an anti-crossing of the QD resonances, which suggests that the direct inter-dot tunnel coupling is negligible compared to processes via the superconductor.
The experiments are performed in a dilution refrigerator at a base-temperature of $\sim20\,$mK.

\begin{figure}[t]
\centering
\includegraphics{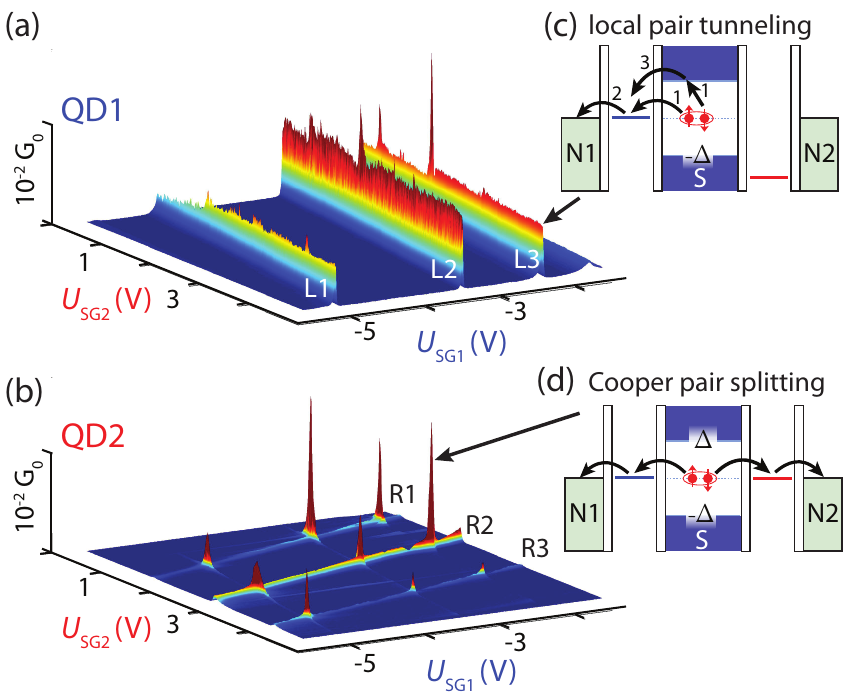}
\caption{(a) Differential conductance of QD1, $G_1$, and (b) of QD2, $G_2$, as a function of the side-gate voltages $U_{\rm SG1}$ and $U_{\rm SG2}$. (c) and (d) Energy diagrams of local pair tunneling (LPT) and CPS.}
\end{figure}

Figures~2a and 2b show the simultaneously recorded differential conductances $G_1$ through QD1 and $G_2$ through QD2, both as a function of the side-gate voltages $U_{\rm SG1}$ and $U_{\rm SG2}$. The measurements were done at zero bias and zero magnetic field.
When $U_{\rm SG1}$ is varied, QD1 is tuned through several resonances, which result in conductance maxima in $G_1$, labeled L1, L2 and L3 in Fig.~2a. The amplitudes of the resonances vary only little when tuning $U_{\rm SG2}$, while the resonance position changes slightly due to capacitive cross talk from SG2 to QD1. Very weak, but similar conductance ridges labeled R1, R2 and R3 can be observed in the conductance through QD2 in Fig.~2b. These are mainly tuned by SG2, which results in conductance ridges almost perpendicular to the ones in Fig.~2a due to QD1.

Our main experimental findings are pronounced peaks when both QDs are in resonance. At these gate configurations the conductance is increased by up to a factor of $\sim 100$ compared to the respective conductance ridge. This is most prominent in $G_2$, but most of the peaks can also be observed in $G_1$ on a larger background. No peaks at resonance crossings can be observed when the superconductivity is suppressed by a small external magnetic field (see below). If only one QD is resonant, only local transport through this QD is allowed. A possible local process is local pair tunneling (LPT), illustrated in Fig.~2c: the first electron of a Cooper pair is emitted into the QD, which leaves S in a virtual excited state. When the first electron has left the dot, the second tunnels into the same QD. Other local processes like double charging of a dot are strongly suppressed by the large charging energies. However, if both QDs are in resonance, the second electron can tunnel into QD2, as shown in Fig.~2d, which splits the initial Cooper pair.

We now focus on the resonance crossing (L2,R2). Figure~3a shows the Coulomb blockade resonance L2 in $G_1$ as a function of $U_{\rm SG1}$ (bottom curve). In the same gate sweep, $G_2$ is tuned through the resonance R2 due to capacitive cross-talk, which results in a wide conductance maximum. However, an additional much sharper peak occurs at the voltage of the L2 resonance, with similar width and shape as the resonance in $G_1$. When the superconductivity is suppressed by an external magnetic field of $250\,$mT, we find no additional peak in $G_2$ at the resonance crossing, but a slight reduction consistent with a classical resistor network \cite{Hofstetter2009}, see inset of Fig.~3a. The resonance positions do not change with field, but the overall conductance can vary strongly due to the superconductor's gap, which reduces local single electron transport and favors LPT.

To assess CPS in the experiments we use the amplitude $\Delta G_2$ of the additional peak in $G_2$ at the position of the QD1 resonance, as illustrated in Fig.~3a. The subtracted background is determined by manually interpolating the bare QD2 resonance. Also indicated in Fig.~3a is the detuning $\delta U$ between the two resonances. Figure~3b shows a series of SG1-sweeps at different values of $U_{\rm SG2}$ near the resonance crossing (L2,R2), with the curve from Fig.~3a highlighted in red. One finds that $\Delta G_2$ depends strongly on the detuning $\delta U$. In Fig.~3c we therefore plot $\Delta G_2$ vs. $\delta U$, where the value of Fig.~3a is marked by a red triangle. As another example, the conductance variation near the crossing (L3,R2) is also plotted in Fig.~3c.
For all crossings we find that $\Delta G_2$ has a maximum at $\delta U\approx0$, i.e. where both QDs are in resonance, in agreement with theoretical predictions \cite{Recher_Loss_PRB63_2001}. For $\delta U\neq 0$, $\Delta G_2$ falls off rapidly and tends to zero on an energy scale consistent with the width of the respective resonances.

On the resonance crossings investigated here the maximum change in $G_2$ is $~0.012\,e^2/h$. This number has to be compared to the total conductance, including the local processes, so that we define the {\it visibility} of CPS in the second branch of the Cooper pair splitter as $\eta_2=\Delta G_2 / G_2$ (similar for $G_1$). The CPS visibilities for both branches on resonance crossing (L3,R2) are plotted in Fig.~3d. $\eta_2$ is essentially constant over a large range of $\delta U$ and reaches values up to 98\%, i.e. the current in one branch can be dominated by CPS. $\eta_1$, however, has a maximum of only $73$\% at $\delta U\approx 0$ and drops to zero for a large detuning.

\begin{figure}[t]
\centering
\includegraphics{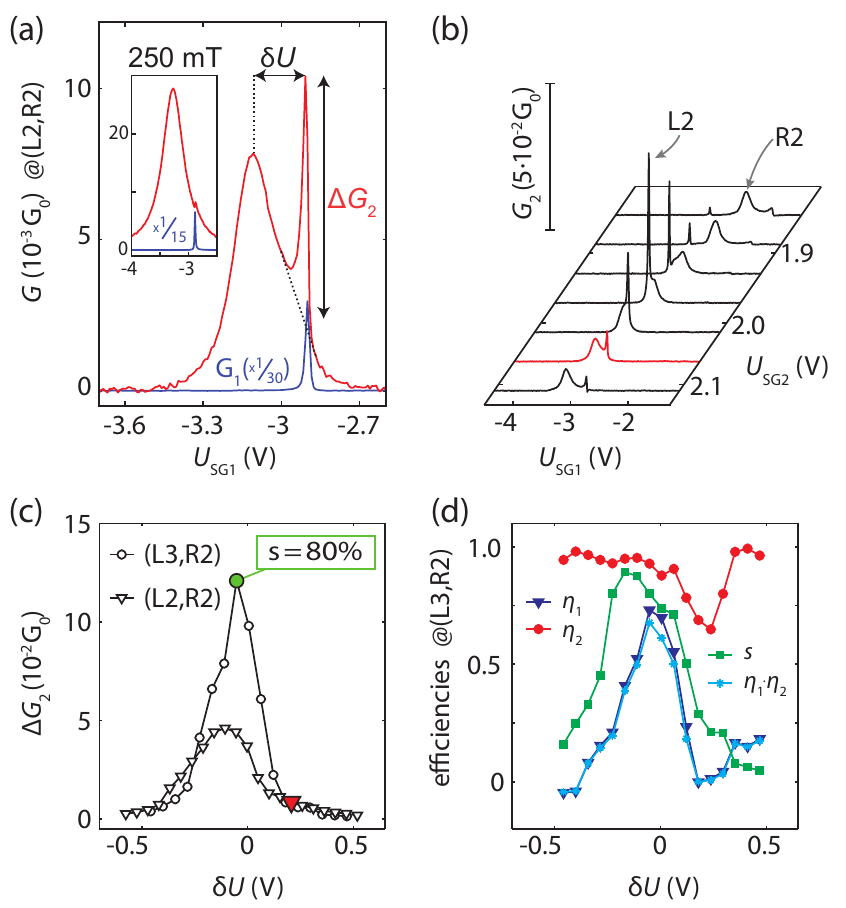}
\caption{(Color online) (a) $G_1$ and $G_2$ as a function of $U_{\rm SG1}$ for $U_{\rm SG2}\approx2.07\,$V. Inset: the same measurement at a magnetic field of $250\,$mT to suppress the superconductivity. (b) $G_2(U_{\rm SG1})$ for a series of side-gate voltages $U_{\rm SG2}$. (c) $\Delta G_2$ as a function of the detuning $\delta U$ between the resonances in QD1 and QD2 for the indicated resonance-crossings. (d) Plots of the visibilities $\eta_i$, the CPS efficiency $s$ and $\eta=\eta_1\eta_2$.}
\end{figure}

As a measure for the CPS efficiency we compare the CPS currents to the total currents in both branches of the device. Assuming that CPS leads to a conductance $G_{\rm CPS}$ in each branch, independent of other processes, we define the CPS efficiency as
\begin{equation}
s=\frac{2G_{\rm CPS}}{G_1+G_2}.
\end{equation}
By setting $G_{\rm CPS}=\Delta G_2$ we find efficiencies up to $s\approx90$\%, much larger than required to demonstrate interaction dominated CPS. The efficiency as a function of $\delta U$ is plotted in Fig.~3d for the crossing (L3,R2). However, depending on the intended purpose of the entangler, $s$ is not necessarily the relevant parameter. For example, in tests of Bell's inequality proposed for electrons \cite{Kawabata_JPSJ_2001, Samuelsson_Buettiker_PRL_2003}, the measured quantities are current cross correlations between the normal metal terminals, which suggests to use the following figure of merit:
\begin{equation}
\eta=\eta_1\cdot\eta_2=\frac{\Delta G_1}{G_1}\cdot\frac{\Delta G_2}{G_2}
\end{equation}
A violation of Bell's inequality requires $\eta >1/\sqrt{2}\approx 71$\%. In Fig.~3d, $\eta$ is plotted as a function of $\delta U$ for the crossing (L3,R2). We find values up to $\eta=68$\%, mostly limited by the large rates of local processes through QD1. Nontheless, the large visibility in $G_2$ demonstrates the feasibility of testing Bell's inequality with electrons, if an ideal detection scheme was available.

Intuitively one might expect $\Delta G_1=\Delta G_2$. This is found within experimental errors for $4$ of the $9$ resonance crossings. As an example, $\Delta G_1$ and $\Delta G_2$ of the crossing (L3,R2) investigated above are plotted as a function of $\delta U$ in Fig.~4a. For the other crossings, the two conductance variations deviate significantly from each other. $4$ of the $9$ crossings exhibit curves comparable to (L3,R1) plotted in Fig.~4b. Here, $\Delta G_2$ is larger than $\Delta G_1$ by about a factor of $2$, but with a similar curve shape. One of the $9$ crossings, (L2,R1) shown in Fig.~4c, is very peculiar: the variation in $\Delta G_1$ is almost negligible, while $\Delta G_2$ exhibits a pronounced peak. In addition, one finds that $\Delta G_2 > G_1$, i.e. the conductance variation in one branch is larger than the total conductance in the other.

To explain our experiments we discuss a strongly simplified semi-classical master equation model. More sophisticated models can be found in \cite{Sauret_Feinberg_Martin_PRB70_2004, Eldridge_Koenig_PRB82_2010}. For each QD we consider a single level with constant broadening and a large charging energy. The system can be in one of the following four states: both QDs empty, either QD filled with one electron, or both dots occupied. The tunneling processes illustrated in Fig.~1a lead to transitions between these states. We assume that effectively electrons are transfered only in one direction, from S to the QDs and from the QDs to the respective normal metal contact. In addition we consider a tunnel coupling between the dots. The $p_{\rm i}$ in Fig.~1a are the probabilities that the corresponding process changes the occupation of a system state. It is not trivial to extract absolute values for the $p_{\rm i}$ from the experiments, especially for the complex transport processes involving S. The resonances are incorporated as a normalized effective density of states. We use a diagrammatic method based on maximal trees to obtain the steady-state occupation probabilities from the corresponding master equation \cite{Schankenberg_RevModPhys_1976}. From the populations of the QDs we then calculate the transport rates.

\begin{figure}[t]
\centering
\includegraphics{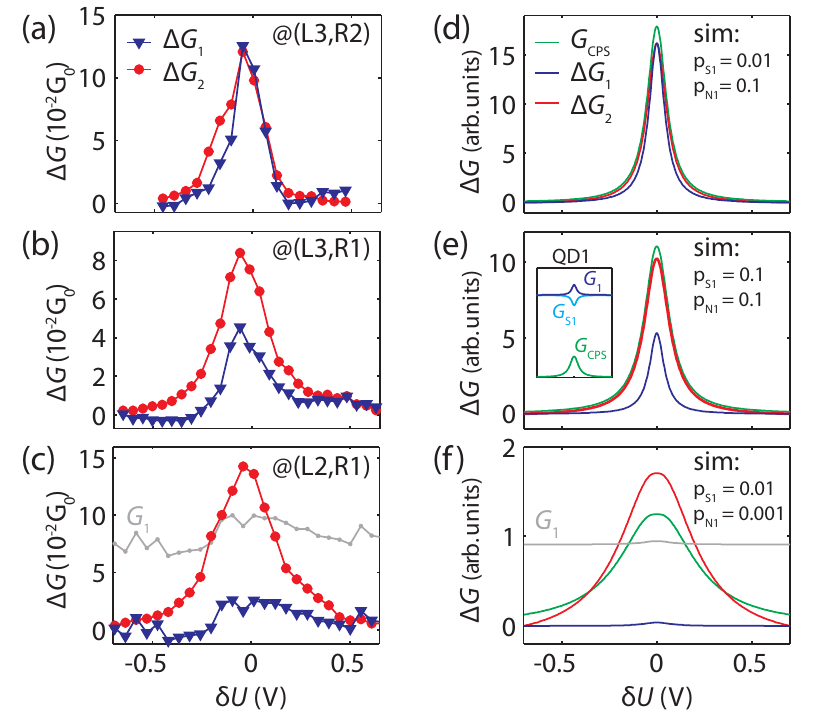}
\caption{(Color online) (a)-(c) $\Delta G_1$ and $\Delta G_2$ as a function of the detuning $\delta U$ for the indicated resonance crossings. (d)-(f) Similar plots obtained from the master equation model of CPS, including $G_{\rm CPS}$. The parameters varied between the simulations are given in the figures. The inset in (e) shows the conductances in the branch of QD1 due to local processes ($G_{\rm S1}$), CPS and at N1.}
\end{figure}

Our model reproduces qualitatively the observed conductance variations and shows that a finite QD population can lead to a competition between the various transport mechanisms. In Figs.~4d-f simulated conductance variations are plotted for different QD1 parameters, while QD2 is kept at $p_{\rm S2}=0.01$ and 
$p_{\rm N2}=0.1$, i.e. in the regime of Ref. \cite{Recher_Loss_PRB63_2001}, where the coupling to S is much weaker than to the normal contacts. We set $p_{\rm CPS}=0.03$ to obtain relative amplitudes comparable to the experiments, and $p_{\rm 12}=0.001$ so that the inter-dot coupling is the smallest parameter in the problem.
If the two current branches are similar, i.e. $p_{\rm S1}\approx p_{\rm S2}<p_{\rm N1} \approx p_{\rm N2}$, one finds $\Delta G_1=\Delta G_2$, as shown in Fig.~4d \cite{footnote}, similar to the experiments presented in Fig.~4a.

For asymmetric branches, the conductance variations are not identical anymore. Figure~4e shows plots for $p_{\rm S1}=p_{\rm N1}=0.1$, for which $\Delta G_2\approx 2\Delta G_1$, as in the experiment shown in Fig.~4b.
The model also allows us to calculate the rate at which Cooper pairs are extracted from S by CPS. The corresponding conductance is also plotted in Figs.~4d-f. We find that $\Delta G_i< G_{\rm CPS}$ as long as the inter-dot coupling $p_{12}$ is negligible, i.e. the experimentally extracted CPS conductance underestimates the actual value, max$(\Delta G_1, \Delta G_2)<G_{\rm CPS}$. The explanation is that due to CPS on a resonance crossing the average QD populations are increased beyond the off-resonance equilibrium due to the local processes, which leads to a reduction of the current into the N contacts. This is illustrated in the inset of Fig.~4e, where the calculated local conductance from S1, $G_{\rm S1}$, has a minimum where $G_{\rm CPS}$ is maximal. Intuitively, the QDs are not emptied fast enough, which inhibits all processes on the dot.

The situation is more complex if the tunnel coupling between the dots becomes relevant. For example, if $p_{\rm N1}=p_{12}=0.001$ and $p_{\rm S1}=0.01$, as used in Fig.~4f, the electrons can leave QD1 to N1 and to QD2 with the same probability. Since $p_{\rm N1}$ is small, this quenches $G_1$, but $G_2$ is increased due to the additional current from QD1. Here, the $\Delta G_i$ do not give an upper or lower bound for the CPS rate and $\Delta G_2$ can become larger than $G_1$, as in the experimental curves in Fig.~4c. We note that the discussed situations are not in the regime of completely filled QDs. Our model suggests that in this unitary limit the conductances can be reduced considerably in the center of a resonance crossing. In our data we find no evidence for this prediction. However, since the coupling between S and an InAs nanowires can be strong, the model might account for the as yet unexplained anomalous behavior of the on-resonance signals in \cite{Hofstetter2009}.

In summary, we present Cooper pair splitting experiments with efficiencies up to 90\%, demonstrating the importance of electron-electron interactions in such systems. For the figure of merit relevant in tests of Bell's inequality for electons we find values close to the required limits. In addition, we asses CPS on both QDs and find rather large apparent discrepancies between the two conductance variations, which we explain in a semi-classical master equation model. The latter suggests that for negligible inter-dot couplings the experimentally extracted CPS rates are a lower bound to the real CPS rates. Our experiments and calculations show that there is a large variety of different transport phenomena in a Cooper pair splitting device that need further investigation. Of capital importance is the observation that if both dots had the properties of QD2, tests of Bell's inequality even with non-ideal detectors could be performed to detect electron entanglement, an imprtant step on the way to a source of entangled electron pairs on demand.

We thank Bernd Braunecker for fruitful discussions and gratefully acknowledge the financial support by the EU FP7 project SE$^2$ND, the EU ERC project QUEST, the Swiss NCCR Nano and NCCR Quantum and the Swiss SNF.

\bibliographystyle{naturemag}

\begin{thebibliography}{1}
\bibitem{footnote}
\bibinfo{note}{For clarity we chose slightly different widths for the QD resonances.}
\end{thebibliography}

\end{document}